\documentclass[aps,preprintnumbers,twocolumn,amsmath,amssymb,prb,showpacs]{revtex4}
\usepackage{bbm}
\usepackage{mathrsfs}
\usepackage{amsmath}
\usepackage{graphicx}
\usepackage{ulem}
\usepackage{color}

\newcommand{\be}{\begin{equation}}
\newcommand{\ee}{\end{equation}}
\newcommand{\bea}{\begin{eqnarray}}
\newcommand{\eea}{\end{eqnarray}}
\newcommand{\bes}{\begin{split}}
\newcommand{\ees}{\end{split}}

\begin{document}
\title{High-Throughput GW Calculations via Machine Learning}
\author{R. A. Abdelghany}
\affiliation{Physics Department, Faculty of Science, Al-Azhar University, Assiut, 71524, Egypt}
\affiliation{Department of Physics, National Chung Hsing University, Taichung, 40227, Taiwan}

\author{Chih-En Hsu}
\affiliation{Department of Physics, Tamkang University, New Taipei 251301, Taiwan}

\author{Hung-Chung Hsueh}
\affiliation{Department of Physics, Tamkang University, New Taipei 251301, Taiwan}

\author{Yuan-Hong Tsai}
\affiliation{AI Foundation, Taipei, 110, Taiwan}

\author{Ming-Chiang Chung}
\email{mingchiangha@nchu.edu.tw}
\affiliation{Department of Physics, National Chung Hsing University, Taichung, 40227, Taiwan}
\affiliation{Physics Division, National Center for Theoretical Sciences, Taipei, 10617, Taiwan}

\begin{abstract}
We present a machine learning (ML) framework that predicts $G_0W_0$ quasiparticle energies across molecular dynamics (MD) trajectories with high accuracy and efficiency. Using only DFT-derived mean-field eigenvalues and exchange-correlation potentials, the model is trained on 25\% of MD snapshots and achieves RMSEs below 0.1 eV. It accurately reproduces k-resolved quasiparticle band structures and density of states, even for BN polymorphs excluded from the training data. This approach bypasses the computational bottlenecks of $G_0W_0$ simulations over dynamic configurations, offering a scalable route to excited-state electronic structure simulations with many-body accuracy.
\end{abstract} 

\date{\today}
\maketitle 

The GW approximation, a cornerstone of many-body perturbation theory, offers a more accurate description of electronic excitations compared to Density Functional Theory (DFT) by correcting the electronic self-energy, particularly addressing DFT's known limitations in predicting band gaps \cite{hedin1965new, golze2019gw, zolyomi2015towards}. This self-energy correction explicitly accounts for electron correlation effects beyond the mean-field approximation, leading to quasiparticle energies that often exhibit excellent agreement with experimental measurements across a broad spectrum of materials. However, a significant computational bottleneck in GW calculations arises from the necessity of computing the inverse dielectric matrix, a process that involves a slowly converging summation over unoccupied states \cite{yeh2022fully, fiedler2022deep, shishkin2006implementation, neuhauser2014breaking}.

Recent investigations have highlighted the crucial role of incorporating molecular dynamics (MD) configurations into GW calculations to achieve a realistic and accurate understanding of temperature- and dynamics-dependent electronic properties \cite{bokdam2016role, tirimbo2020excited, grobas2021time, perfetto2022real, zacharias2016one, zacharias2020theory, segalina2021structure, wiktor2017predictive, leppert2024excitons, manjanath2024non}. For instance, in halide perovskites, temperature-induced variations in metal-halide bond angles and lengths have a substantial impact on their band gaps, underscoring the necessity of combining MD simulations with GW calculations \cite{wiktor2017predictive}. Similarly, temperature-dependent modifications to the electronic structure significantly influence optical absorption and emission spectra, which are critical parameters for optoelectronic and photovoltaic applications \cite{zacharias2016one, zacharias2020theory}. MD simulations provide snapshots of atomic configurations that inherently capture dynamic disorder at finite temperatures, offering a more physically relevant description compared to static calculations \cite{zacharias2016one, wiktor2017predictive, zacharias2020theory, segalina2021structure}. Nevertheless, the computational cost associated with performing GW calculations for a statistically significant number of MD configurations remains a substantial challenge \cite{leppert2024excitons, wiktor2017predictive, segalina2021structure, manjanath2024non}.

Machine learning (ML) provides a promising avenue for accelerating GW computations by exploiting its capacity to learn complex patterns from data \cite{rasmussen2021towards, knosgaard2022representing, hou2024unsupervised, zauchner2023accelerating, caylak2021machine, lee2016prediction, ghosh2024predicting, jung2024automatic}. Previous studies have explored a variety of strategies to predict material properties, utilizing features ranging from elemental attributes and atomic descriptors to structural characteristics such as Kohn-Sham band gaps, cohesive energy, and crystalline volume \cite{knosgaard2022representing, lee2016prediction, hou2024unsupervised, caylak2021machine}. However, many of these approaches are constrained to predicting scalar properties, such as band gaps, rather than capturing the full k-resolved band structure \cite{lee2016prediction, ghosh2024predicting, jung2024automatic}. Moreover, constructing features that encapsulate all the relevant information about electronic states and their interactions remains a significant challenge. This process often involves extensive manual feature engineering, which can produce unphysical outcomes requiring post-processing, and the reliance on human intuition for feature selection can introduce bias \cite{knosgaard2022representing, hou2024unsupervised}.

In this Letter, we introduce an efficient ML-based method for predicting GW quasiparticle energies using a minimal set of features derived from DFT. Utilizing 1000 molecular dynamics (MD) snapshots each for BN and Si, our model is trained on only 25\% of the data and accurately predicts quasiparticle energies for the remaining 75 \% using only readily accessible DFT quantities such as mean-field eigenvalues and exchange-correlation potentials. In contrast to prior approaches that depend on handcrafted descriptors, our method streamlines the workflow by eliminating additional feature engineering, while preserving both high accuracy and interpretability. Crucially, the model generalizes to unseen BN polymorphs with minimal loss in accuracy, demonstrating its transferability and robustness across structural variations.

\begin{figure}[ht]
\includegraphics[width=\columnwidth]{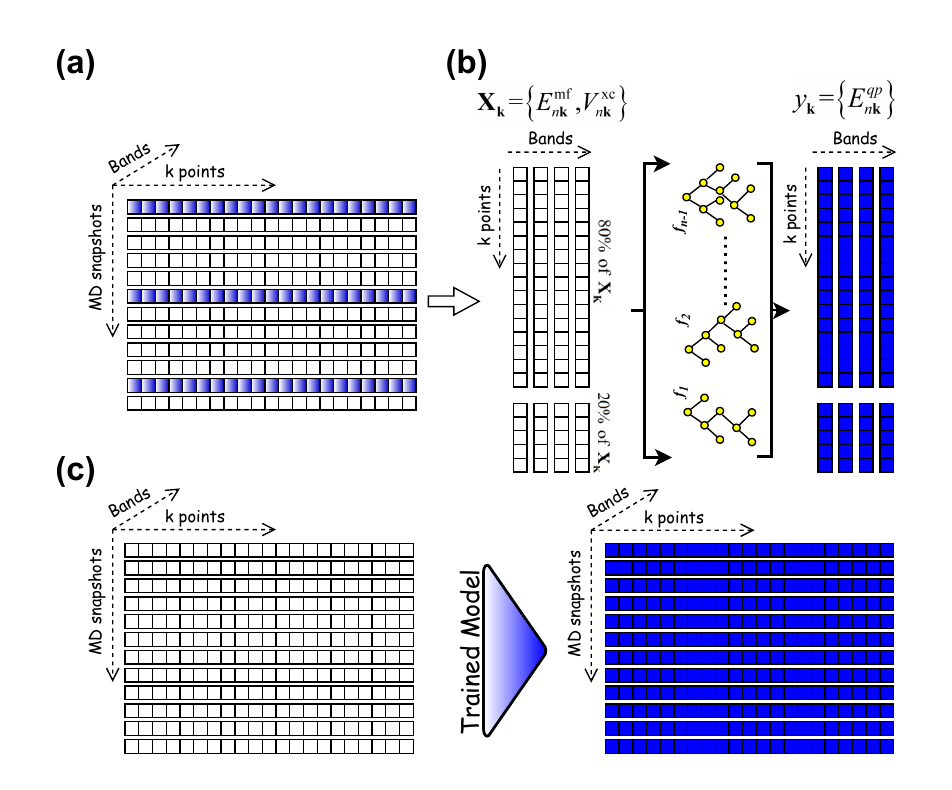}
\caption{\label{fig: ML_model} Schematic representation of the proposed machine learning model for predicting GW quasiparticle energies. \textbf{(a)} shows a 3D grid representing the input data structure, where rows correspond to MD snapshots, columns to k-points, and depth to energy bands. Partially blue-colored rows indicate to the selected snapshots used in training the model. \textbf{(b)} illustrates the feature engineering and training processes, where ${{\bf{X}}_{\bf{K}}}$ are partitioned into 80$\% $ for training and 20$\% $ for testing. The lower panel, \textbf{(c)} demonstrates the application of the trained model to predict GW quasiparticle energies across all MD snapshots.}
\end{figure}

Our computational workflow is illustrated in Fig. \ref{fig: ML_model}. We begin with a series of MD simulations, generating a diverse set of structural snapshots over time. A subset of these snapshots, 25$\% $ of all snapshots, (partially blue-colored rows), strategically selected to represent the entire simulation trajectory, is used for training the machine learning model. The feature engineering process, detailed in the upper-right panel, involves constructing input features by concatenating the mean-field energies and the exchange-correlation potential, both truncated to a specified number of energy bands ${{\bf{X}}_{\bf{K}}} = \left\{ {E_{n{\bf{K}}}^{mf},V_{n{\bf{K}}}^{xc}} \right\}$. These quantities, capturing aspects of the local electronic environment and single-particle energies, are combined into feature vectors for each k-point within each training snapshot. These feature vectors, along with their corresponding GW quasiparticle energies ${y_{\bf{K}}} = \left\{ {E_{n{\bf{K}}}^{qp}} \right\}$, also truncated to the same number of bands, are then aggregated to form the training (80\%) and validating (20\%) datasets. Finally, the trained model is used to predict the GW quasiparticle energies for all MD snapshots, as shown in the lower panel.

A gradient-boosting machine learning model, specifically a LightGBM regressor \cite{ke2017lightgbm}, was employed within a multi-output regression framework to predict the GW quasiparticle energies. This approach allows us to simultaneously predict multiple energy levels associated with each k-point. 
Within the model, a series of intermediate functions \( f_1, f_2, \ldots, f_{n-1} \) sequentially transform the input features. Each \( f_i \) corresponds to a decision tree in the gradient-boosting framework, where the output of \( f_i \) is passed to \( f_{i+1} \), incrementally refining the prediction by minimizing the loss function. This hierarchical structure captures the complex, non-linear relationships between the mean-field and exchange-correlation terms and the target quasiparticle energies.
The LightGBM hyperparameters was optimized using Optuna \cite{akiba2019optuna}, a hyperparameter optimization framework, to maximize the average R-squared score across the predicted energy bands. Full details regarding model architecture, training procedure, and hyperparameter optimization are provided in the supplementary material.

This study utilizes silicon (Si) and boron nitride (BN) as representative materials. Both materials were initially relaxed structures obtained from the Materials Project database \cite{jain2013commentary}. Specifically, Si corresponds to Materials Project ID mp-149 (space group $Fd\bar 3m$, band gap 0.61 eV), and BN corresponds to Materials Project ID mp-2653 (space group $P6_3mc$ , band gap 5.2 eV) . Car-Parrinello molecular dynamics (MD) simulations were then performed for each material at 300 K using the NVT ensemble in QUANTUM ESPRESSO \cite{giannozzi2009quantum, giannozzi2017advanced, giannozzi2020quantum}. Following equilibration, 1000 MD configurations were extracted for both Si and BN, equally spaced across the simulation trajectories to create a representative dataset. GW calculations were then performed on all snapshots using BerkeleyGW \cite{hybertsen1986electron, deslippe2012berkeleygw}, employing 150 Kohn-Sham bands to ensure convergence of quasiparticle energies. Thermal fluctuations from MD simulations induce structural variations that modify the electronic structure, including band gap shrinkage and transient metallization in some configurations (see Supplemental Material for details) \cite{zhang2012quantum, liu2017two, takai2017size, lu2018hittorf, ye2018germanene, konstantinou2018origin, mei2019structural, nhan2023density, dong2022tunable, hu2024molecular}.

\begin{figure}[ht]
	\centering
	\begin{minipage}{0.49\columnwidth}
		\centering
		\textbf{(a)}\\
		\includegraphics[width=\textwidth]{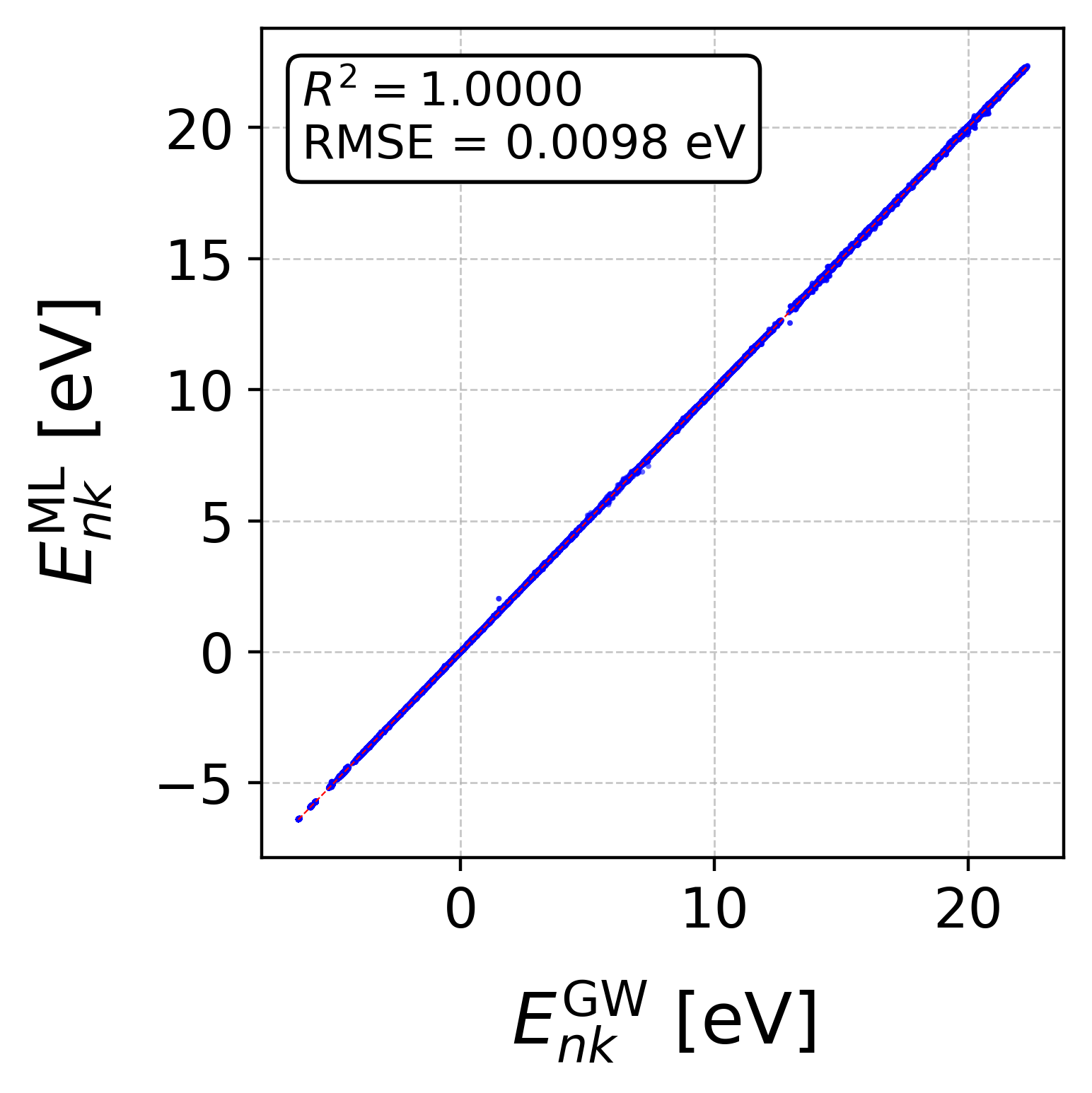}
	\end{minipage}
	\hfill
	\begin{minipage}{0.49\columnwidth}
		\centering
		\textbf{(b)}\\
		\includegraphics[width=\textwidth]{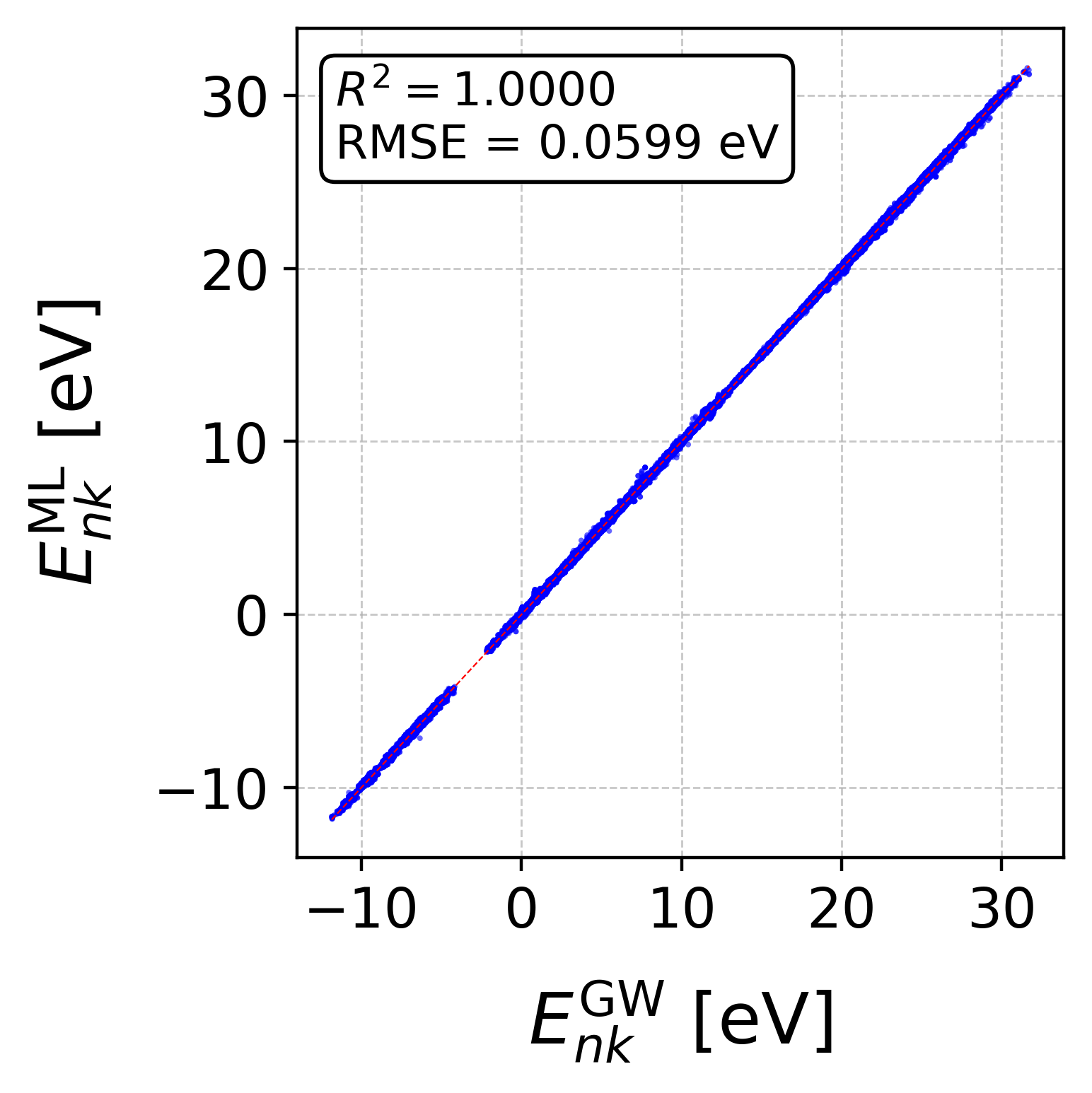}
	\end{minipage}
	
	\vspace{1em} 
	
	\begin{minipage}{0.49\columnwidth}
		\centering
		\textbf{(c)}\\
		\includegraphics[width=\textwidth]{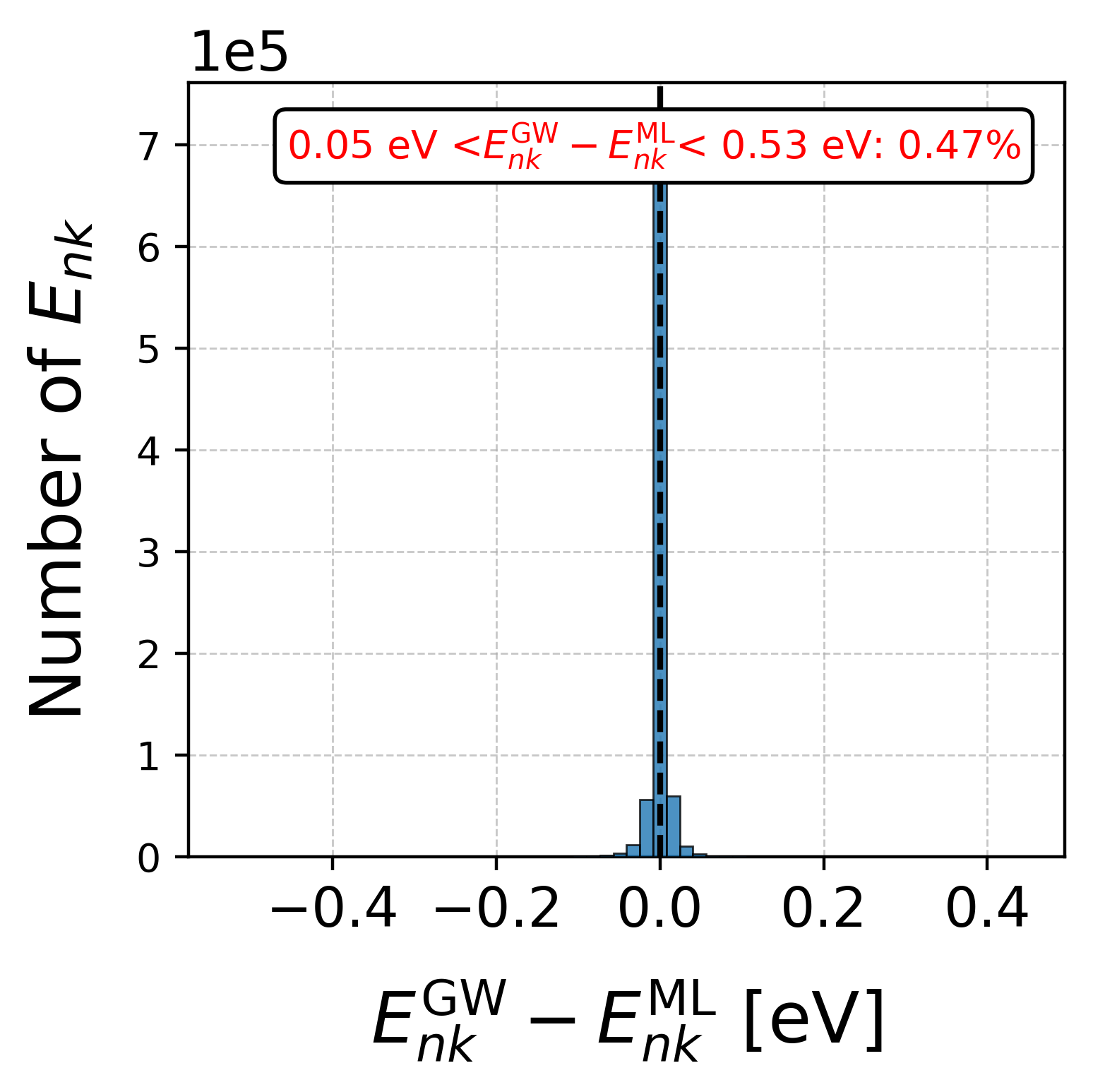}
	\end{minipage}
	\hfill
	\begin{minipage}{0.49\columnwidth}
		\centering
		\textbf{(d)}\\
		\includegraphics[width=\textwidth]{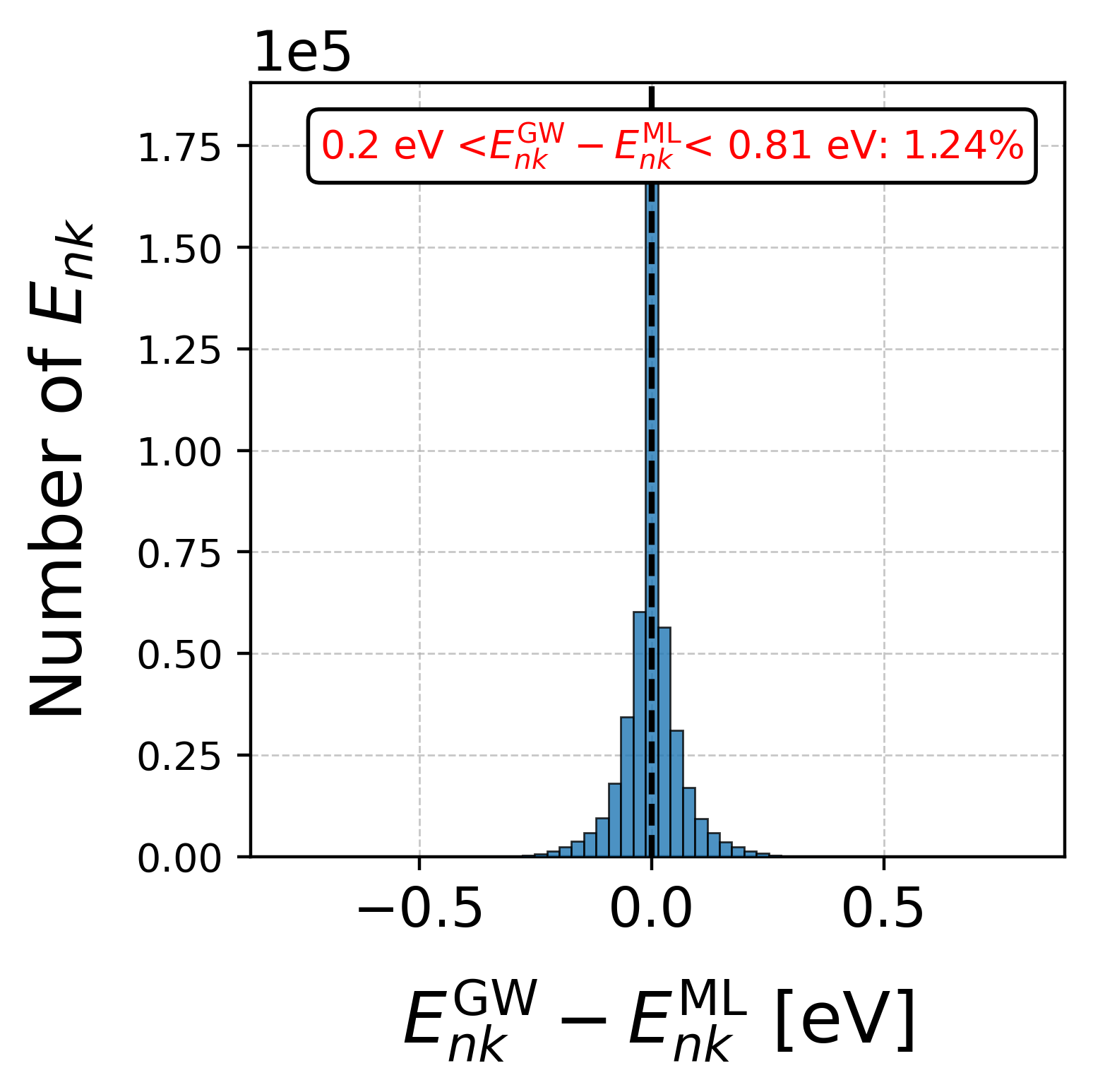}
	\end{minipage}
	
	\caption{Predicted quasiparticle energies for all MD snapshots of Si(mp-149) (left panel) and BN(mp-2653) (right panel). Figs. (a) and (b) present parity plots comparing the predicted quasiparticle energies to the reference GW values, with the red dashed line representing ideal predictions. Figs. (c) and (d) display histograms illustrating the distribution of prediction errors.}
	\label{fig:predis_1}
\end{figure}

Separate machine learning models were trained for each material: Si (mp-149) and BN (mp-2653). For each material, 25\% of the MD snapshots were used for training and validation, with the remaining 75\% reserved for testing the model's ability to generalize. Fig. \ref{fig:predis_1} presents the accuracy of the trained models with the testing datasets, including parity plots of predicted vs. calculated GW quasiparticle energies and histograms of prediction residuals. The root-mean-square error (RMSE) for the Si model was 0.0098 eV ($R^2 \approx $ 1.0), and for the BN model, the RMSE was 0.0599 eV ($R^2 \approx $ 1.0). These high $R^2$ values indicate a strong correlation between predicted and calculated quasiparticle energies, signifying excellent model performance. The histograms of prediction residuals show narrowly peaked, symmetric distributions around zero, suggesting minimal systematic bias. Specifically, for Si, only 0.47$\%$ of predictions had absolute errors between 0.05 eV and the maximum observed error of 0.53 eV. For BN, 1.24$\%$ of the errors were between 0.2 eV and 0.81 eV. This indicates that the majority of prediction errors are concentrated within a narrow range of $\pm$0.2 eV, demonstrating the model's ability to accurately capture the relationship between input features and GW quasiparticle energies.

\begin{figure}[ht]
	\centering
	\begin{minipage}{0.49\columnwidth}
		\centering
		\textbf{(a)}\\
		\includegraphics[width=\textwidth]{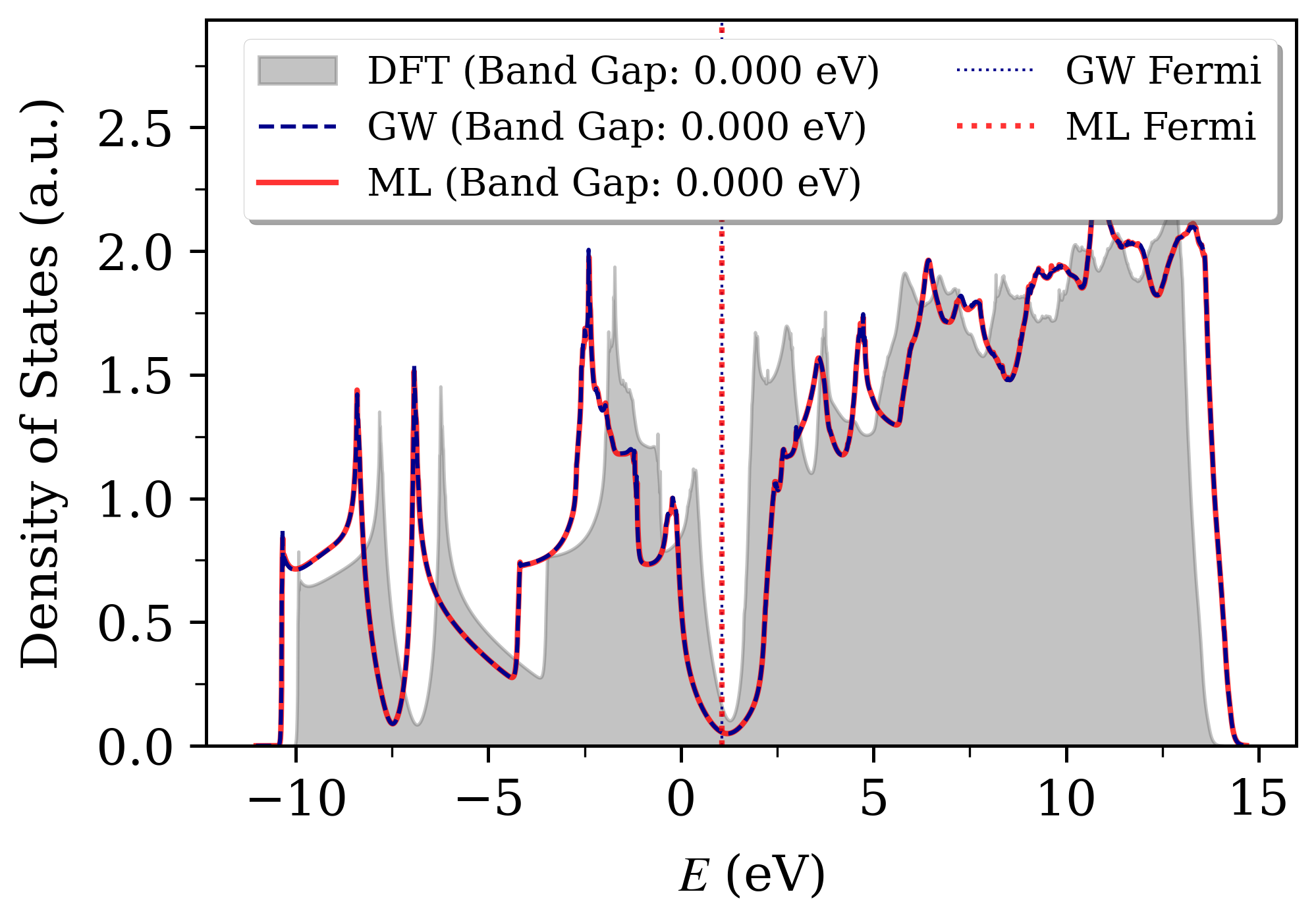}
	\end{minipage}
	\hfill
	\begin{minipage}{0.49\columnwidth}
		\centering
		\textbf{(b)}\\
		\includegraphics[width=\textwidth]{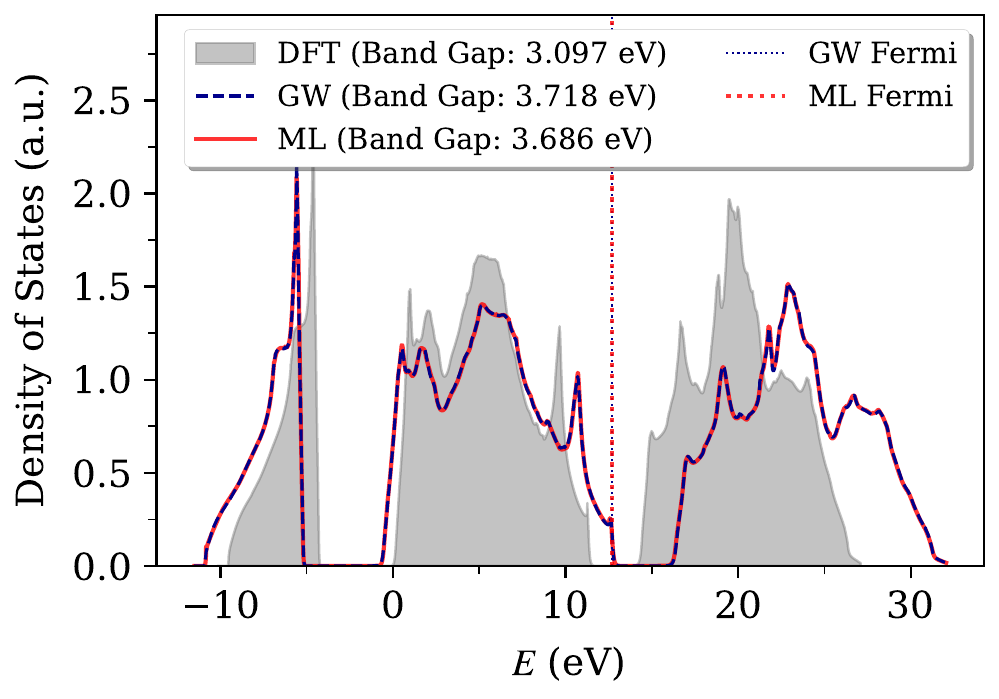}
	\end{minipage}
	
	\vspace{1em} 
	
	\begin{minipage}{0.49\columnwidth}
		\centering
		\textbf{(c)}\\
		\includegraphics[width=\textwidth]{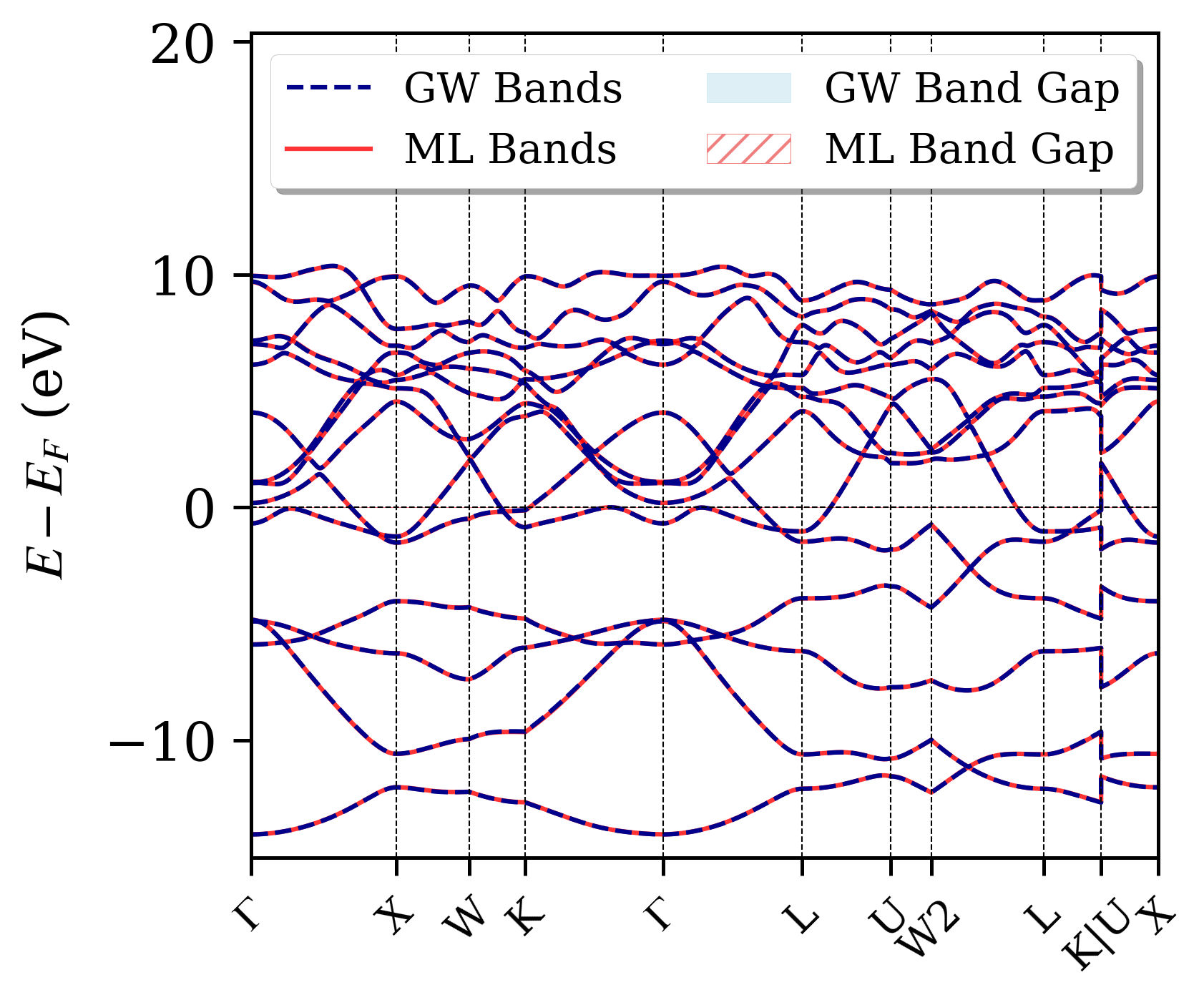}
	\end{minipage}
	\hfill
	\begin{minipage}{0.49\columnwidth}
		\centering
		\textbf{(d)}\\
		\includegraphics[width=\textwidth]{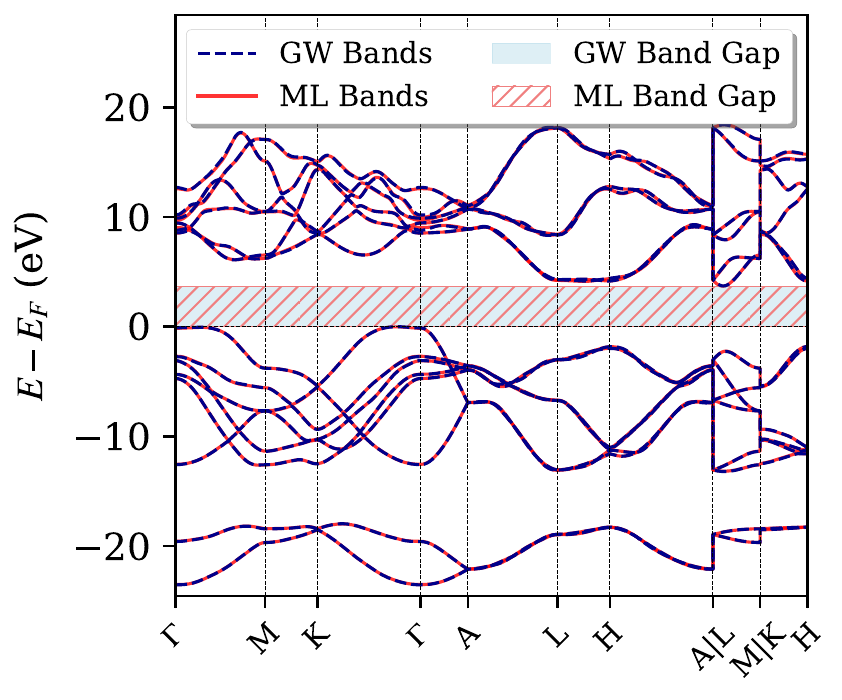}
	\end{minipage}
	
	\caption{DOS and band structures for arbitrarily selected snapshot. The left panel corresponds to Si(mp-149) and the right panel to BN(mp-2653). Corresponding DFT-calculated DOS are included for comparison.}
	\label{fig:DOS_BS}
\end{figure}

The model's ability to predict k-resolved electronic structure is demonstrated in Fig. \ref{fig:DOS_BS}, which presents predicted density of states (DOS) and band structures for arbitrarily selected snapshots, alongside corresponding DFT-calculated DOS for comparison. As shown, the predicted DOS and band structures closely match the reference GW calculations across both valence and conduction band regions. Notably, accurate predictions for 1000 MD snapshots were achieved by training the model with only 250 snapshots, highlighting the efficiency of our approach. In contrast, recent studies predicting only HOMO and LUMO energies required significantly larger datasets (e.g., 6000 snapshots for water monomers/dimers, 10,000 for acetone in water) and complex features including Cartesian coordinate-based representations, orbital-sensitive representations (OSRs), and Kohn-Sham orbital energies \cite{caylak2021machine}.

For BN (mp-2653), the arbitrarily chosen snapshot exhibits a GW-calculated band gap of 3.718 eV, while the predicted band gap is 3.686 eV, corresponding to a deviation of $\Delta E = 0.032$ eV. This agreement validates the accuracy of the model in capturing the electronic structure features for both occupied and unoccupied states.

To systematically examine the correlation between model performance and the size of the training dataset, $R^2$ and MSE were evaluated for the hold-out dataset as a function of the percentage of data utilized for training, as illustrated in Fig.S2 in the supplementary material. The $R^2$ value increases rapidly with the training data size, reaching a plateau beyond approximately 25$\%$. Conversely, the MSE exhibits an inverse trend, decreasing from 0.176 eV when the training data is at its smallest size (7$\%$) and continuing to decrease as the training dataset expands.

\begin{figure}[ht]
	\centering
	\begin{minipage}{0.49\columnwidth}
		\centering
		\textbf{(a)}\\
		\includegraphics[width=\textwidth]{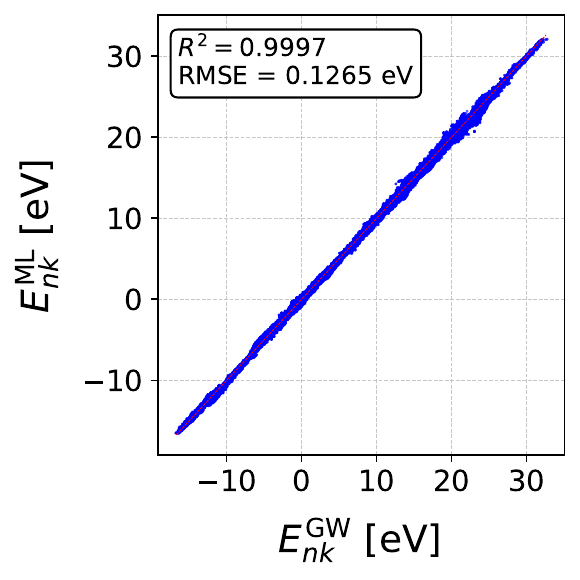}
	\end{minipage}
	\hfill
	\begin{minipage}{0.49\columnwidth}
		\centering
		\textbf{(b)}\\
		\includegraphics[width=\textwidth]{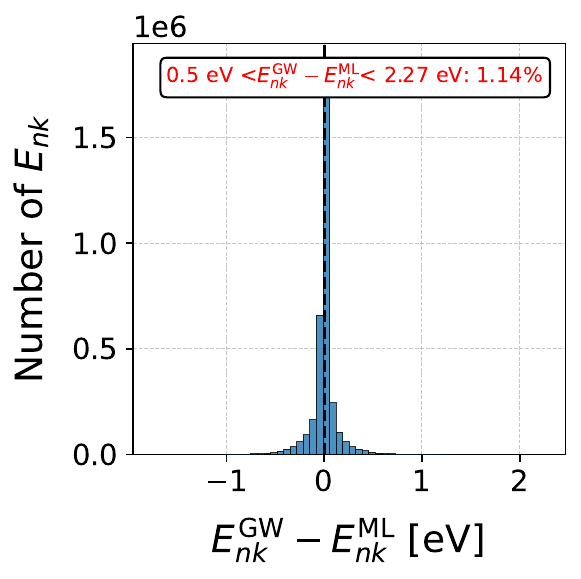}
	\end{minipage}
	
	\vspace{1em} 
	
	\begin{minipage}{0.99\columnwidth}
		\centering
		\textbf{(c)}\\
		\includegraphics[width=\textwidth]{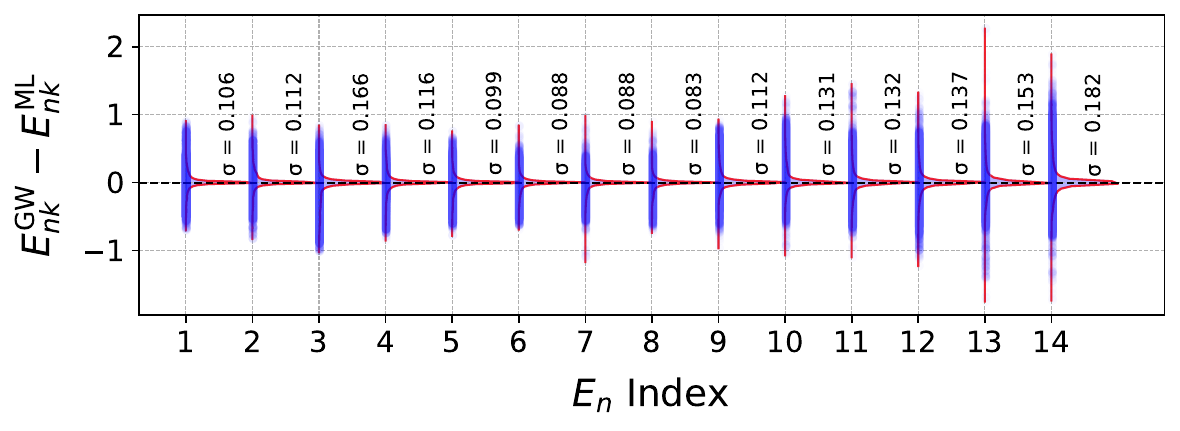}
	\end{minipage}
	
	\caption{Predicted quasiparticle energies for BN polymorphs: BN(mp-601223), BN(mp-2653), BN(mp-7991), and BN(mp-604884). (a) Parity plot comparing predicted quasiparticle energies to reference GW values, with the red dashed line representing ideal predictions. (b) Histogram of prediction errors. (c) Band-resolved analysis of prediction errors in GW quasiparticle energies, with red Gaussian curves representing the error distribution for each band.}
	\label{fig:predis_2}
\end{figure}

\begin{figure}[ht]
	\centering
	\begin{minipage}{0.49\columnwidth}
		\centering
		\textbf{(a)}\\
		\includegraphics[width=\textwidth]{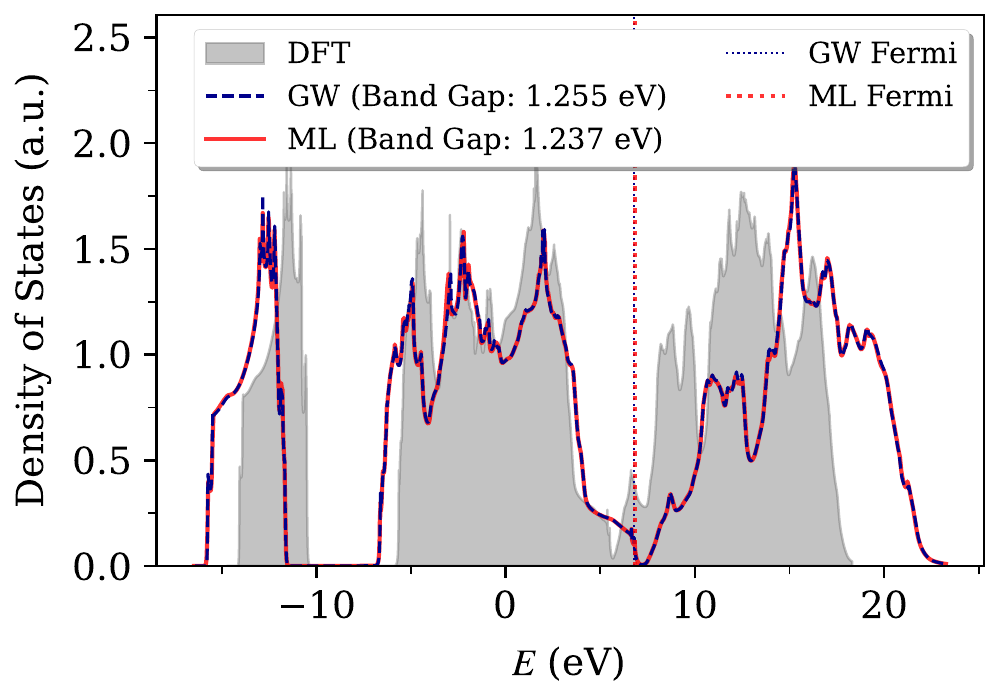}
	\end{minipage}
	\hfill
	\begin{minipage}{0.49\columnwidth}
		\centering
		\textbf{(b)}\\
		\includegraphics[width=\textwidth]{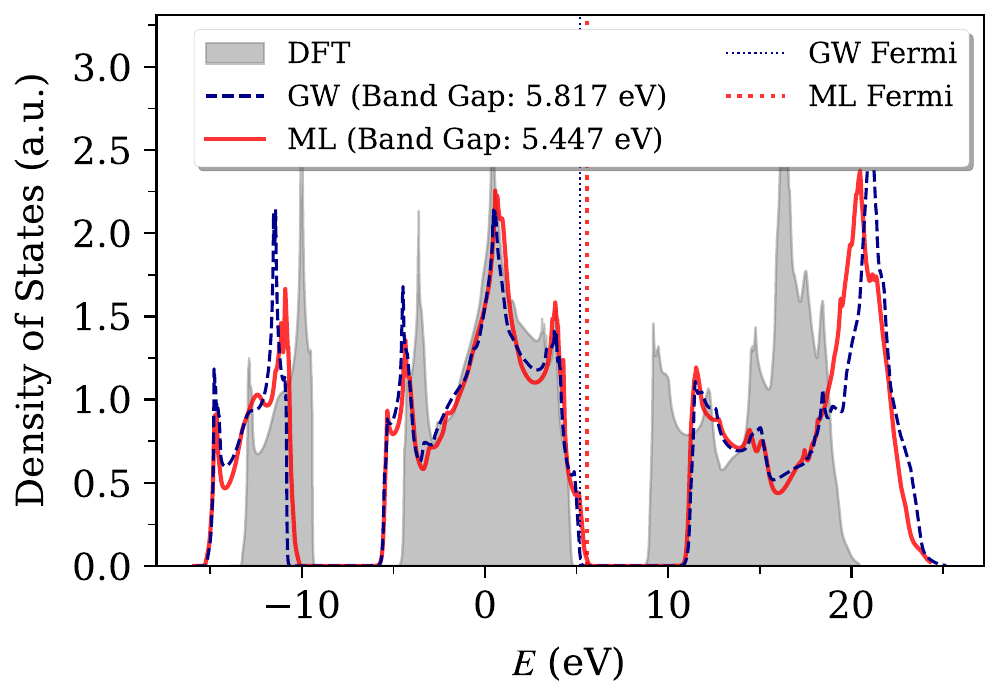}
	\end{minipage}
	
	\vspace{1em} 
	
	\begin{minipage}{0.49\columnwidth}
		\centering
		\textbf{(c)}\\
		\includegraphics[width=\textwidth]{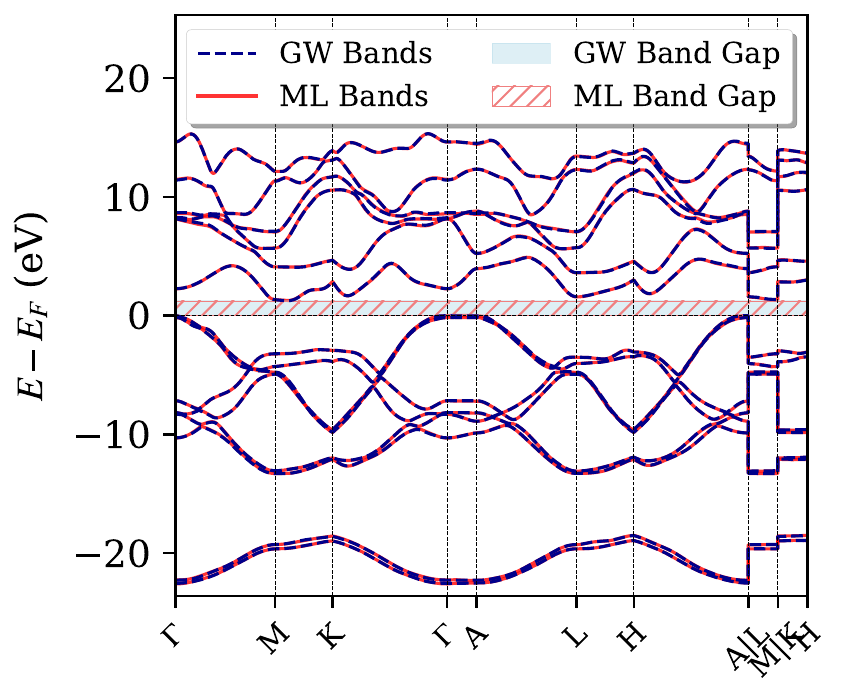}
	\end{minipage}
	\hfill
	\begin{minipage}{0.49\columnwidth}
		\centering
		\textbf{(d)}\\
		\includegraphics[width=\textwidth]{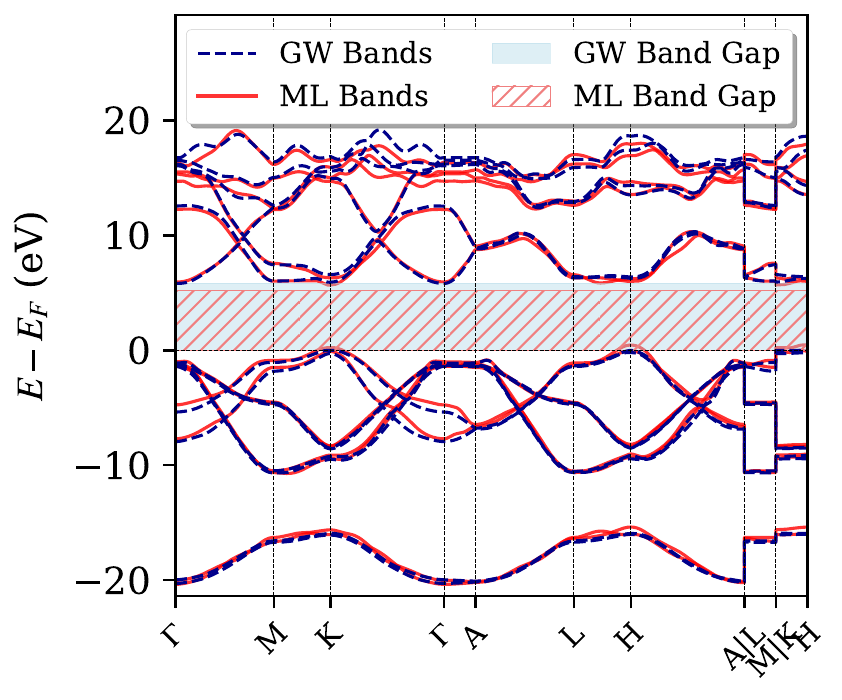}
	\end{minipage}  
	
	\caption{DOS and k-resolved band structure, demonstrating the ability of the model to accurately reproduce electronic properties despite being trained on diverse polymorphic structures. The left panel is for an unseen snapshot of BN, and the right panel is for an unseen BN polymorph structure, BN(mp-604884). The model predicts a band gap of 5.45 eV for BN(mp-604884), with a deviation of $\Delta E$ = 0.37 eV from the calculated GW band gap of 5.82 eV, highlighting its accuracy and generalizability to untrained BN structures.}
	\label{fig:DOS_BS2}
\end{figure}

To further test model robustness and capture greater structural diversity, the training dataset was enriched by combining MD simulation data for BN (mp-2653) with data from additional polymorphs, BN (mp-601223) and BN (mp-7991), which have band gaps of 0 and 3.96 eV, respectively. The trained model was then used to predict the quasiparticle energies of a new polymorph, BN (mp-604884, space group $p\bar6m2$, band gap 4.37 eV). This inclusion of multiple boron nitride polymorphs underscores the robustness and generalizability of our machine learning approach. Using only 25$\%$ of the available data for training, the model accurately captured structural variations observed during MD simulations and predicted quasiparticle energies for the unseen BN polymorph (mp-604884), as demonstrated in Fig. (\ref{fig:predis_2}) and Fig. (\ref{fig:DOS_BS2}). This capability enables reliable predictions of quasiparticle energies across diverse structural environments. The model's transferability and predictive accuracy for previously unseen polymorphs offer a cost-effective and efficient alternative to computationally intensive first-principles calculations for exploring the electronic properties of BN materials.

Fig. \ref{fig:predis_2}a shows a parity plot comparing predicted quasiparticle energies with reference values for the 75\% of snapshots excluded from training and the MD snapshots of BN (mp-604884), which were entirely unseen by the model during training. The predictions closely align with the ideal reference line (red dashed line), yielding an RMSE of 0.126 eV and a high coefficient of determination ($R^2$ = 0.9997). The minimal spread around the ideal line indicates high predictive accuracy. The histogram of prediction errors, Fig. \ref{fig:predis_2}a, centered near zero, confirms minimal bias, while the narrow, sharply peaked distribution indicates low variance and high precision. Only 1.14$\%$ of the dataset exhibits errors ranging from 0.5 eV to 2.27 eV, suggesting robust performance across the majority of the dataset. However, when the model was trained using only the first 25\% of the MD snapshots, rather than an equally distributed subset,the performance significantly deteriorated, with the RMSE increasing to 0.257 eV. This highlights the importance of sampling diversity across the MD trajectory to capture the full range of structural variations and ensure accurate quasiparticle energy predictions.

To identify bands contributing to larger discrepancies, a band-resolved analysis of prediction errors was conducted, considering the 14 electronic bands included in the calculations. Fig. 4c illustrates the error distribution ($E_{nk}^{\mathrm{GW}} - E_{nk}^{\mathrm{ML}}$) for each band, with red Gaussian curves representing these distributions. The standard deviation ($\sigma $) reveals that the largest errors occur in predictions for the outermost conduction bands, while bands near the HOMO and LUMO exhibit smaller errors. This trend is consistent with the expectation that higher-energy conduction bands, being less populated and more sensitive to electronic correlations, are more challenging to predict accurately. The model demonstrates greater reliability for bands closer to the Fermi level, which are typically of greater practical interest.

Fig. \ref{fig:DOS_BS2} demonstrates the model's ability to predict k-resolved band structure and DOS for both unseen MD snapshots (left panel) and the unseen BN polymorph BN(mp-604884) (right panel). The left panel illustrates accurate reproduction of the k-resolved band structure and DOS for representative snapshots, despite the model being trained on diverse polymorphic structures of BN, highlighting its generalizability. The right panel showcases predictive power by simulating key features of the DOS for BN(mp-604884). The model predicts a band gap of 5.45 eV, deviating by only  $\Delta E$ = 0.37 eV from the calculated GW band gap of 5.82 eV. This close agreement underscores the ability to capture essential electronic properties of novel BN structures. Across all MD snapshots of BN(mp-604884), the model demonstrates an RMSE of 0.281 eV for band gap predictions. This low RMSE indicates that the model effectively captures electronic structure variations along the MD trajectories, even for a polymorph not included in the training dataset. These results emphasize the model's potential as a reliable tool for rapid and accurate electronic structure predictions, facilitating the exploration of new materials with minimal computational cost.

Traditional GW quasiparticle energy calculations require both Sigma and Epsilon computations, which are highly time-consuming. For a single MD snapshot, Sigma calculations alone can take approximately 3 hours of wall time, and this time increases dramatically with the number of bands considered in the mean-field calculations, making large-scale simulations impractical. In contrast, our machine learning model predicts GW quasiparticle energies for all snapshots within seconds, eliminating the need for costly Epsilon calculations. This approach drastically reduces computational expense while maintaining accuracy, demonstrating the potential of ML-driven methods in electronic structure calculations.

To evaluate the accuracy of our machine learning model, we compare its performance on predicting the band gap of BN(mp-7991) MD snapshots, Fig. 4S in Supplemental Material,  with several recent studies in the literature. Knøsgaard et al. \cite{knosgaard2022representing} trained a model using DFT-based electronic fingerprints to predict $G_0W_0$ quasiparticle energies across a dataset of 286 2D semiconductors. Their reported MAEs for band gaps ranged from 0.15 to 0.23 eV, and RMSEs ranged from 0.21 to 0.26 eV, depending on the selection of training bands and inclusion of polarizability. Hou et al. \cite{hou2024unsupervised} employed a variational autoencoder (VAE) to learn DFT wavefunction representations and used a neural network to predict k-resolved GW corrections, achieving an MAE of 0.11 eV for individual quasiparticle energy states across a test set of 2D materials. Zauchner et al. \cite{zauchner2023accelerating} developed a machine learning method to predict density-density response functions (DDRFs) for accelerating GW calculations on hydrogenated silicon clusters. Their ML-GW approach yielded RMSEs of 0.06–0.15 eV for HOMO-LUMO gaps and errors in quasiparticle shifts generally within 0.1 eV when compared to GW calculations using the same basis set. Our model, focused on MD snapshots of a single material, achieved an RMSE of just 0.043 eV for the band gap of BN(mp-7991), which is notably lower than the band gap and state-level errors reported in these broader studies. Although our model benefits from reduced material diversity, the results highlight its high precision in capturing quasiparticle energy variations under thermal fluctuations.


In conclusion, given the growing recognition of the critical role of non-equilibrium states and excited-state dynamics in materials science, as evidenced by recent studies employing the GW approach, this work introduces a streamlined machine learning framework for the efficient prediction of GW quasiparticle energies. By obviating the need for computationally demanding GW calculations across all molecular dynamics (MD) snapshots, the proposed method achieves high accuracy with a training set comprising only 25\% of the available data. Notably, the model demonstrates robust generalizability by accurately predicting the density of states (DOS) and k-resolved band structure of boron nitride polymorphs entirely excluded from the training phase. This efficiency stands in stark contrast to contemporary machine learning studies that often require training datasets encompassing 80-90\% of the data and frequently necessitate the computation of numerous auxiliary input features. This study thus presents a computationally tractable pathway to achieving GW-level accuracy in MD simulations, thereby enabling the exploration of intricate electronic phenomena in dynamically evolving environments.


\begin{acknowledgments}
Ming-Chiang Chung acknowledges the NSTC support under Contract
No. 113-2112-M-005-002- and the Asian Office of Aerospace Research and Development (AOARD) for support under Award No. FA2386-23-1-4104. Hung-Chung Hsueh like to thank the National Science and Technology Council
(NSTC) of Taiwan for providing financial support for research under the Projects No. 113-2112-M-032-013. We also thank to National Center for High-performance Computing (NCHC) of National Applied Research Laboratories (NARLabs) in Taiwan for providing computational and storage
resources.
\end{acknowledgments}

\end{document}